\documentclass[prb,amsmath,floatfix,showpacs,twocolumn]{revtex4}
\usepackage{graphicx}
\usepackage{bm}
\def\be{\begin{equation}}
\begin{document}
\title{Flux transitions in a superconducting ring}
\author{Jorge Berger}
\affiliation{Physics Unit, Ort Braude College, P. O. Box 78,
21982 Karmiel, Israel and \\
Department of Physics, Technion, 32000 Haifa, Israel}
\email{phr76jb@tx.technion.ac.il}
\begin{abstract}
We consider a superconducting ring embedded in a magnetic field. The
magnetic field is initially too high to support a superconducting state, but 
is ideally slowly decreased and then reversed. During this process, the number 
of fluxoids trapped by the ring exhibits a complex pattern of transitions,
which depends qualitatively on the ratios of the geometric dimensions of the 
ring to the coherence length. We evaluate the current around the ring 
and find rough agreement with experiments. However, we
also predict a new feature that has not shown up in experiments: in the
vicinity of a critical field, giant jumps are expected.
\end{abstract}
\maketitle

   
\section{\label{INT}INTRODUCTION}
It has been long known that a superconducting loop embedded in a magnetic field exhibits
phenomena of fundamental nature, such as persistent currents and flux quantization.\cite{q1,q2}
The concept of flux quantization has been crucial for the analysis of a wide range of topics of
basic\cite{LP,AB} and practical\cite{jos} importance. The ability to manufacture loops
with widths of the order of the coherence length brought renewed interest in this system;
experimental\cite{mosh,Liu} and theoretical\cite{CS,Denis} studies are being performed
until present.

The purpose of the present article is the theoretical study of the following
situation. A ring of superconducting material is embedded in a homogeneous 
magnetic field. Initially the field is too high to support a superconducting
state, but it is slowly lowered to zero while the temperature is
kept constant; the field is then reversed and
increased until superconductivity is destroyed.
The most prominent feature of these experiments comes from the thermodynamic
need for a winding number (number of fluxoids) which is not too far from the
applied flux,\cite{tink} together with the constraint that this number has to be integer.
These coupled requirements usually lead to jumps in the winding number and,
accordingly, in quantities such as the current around the ring. 

Experiments like this were performed long ago, but modern versions of it are
still giving new results.\cite{Price,Kanda,Ped,GM} 
Situations similar to that considered here have been
reported in Refs.~\onlinecite{Pann,Baelus}.
In the limit of a thin ring, this situation provides a neat example for the
study of state selection in phase transitions.\cite{T1,T2}
Similar experiments for
weakly connected rings,\cite{Silver} and for disks\cite{Geim,Fink,Peeters} 
have also been widely studied.

In the limit of infinite radius, a ring becomes a slab with parallel plane
boundaries. We shall suggest
in Sec.~\ref{onset} that the most outstanding feature obtained in our results
can be related to the condition for the formation of vortices in
a slab parallel to the magnetic field.\cite{SJ,book,Schul,Fink1} If the coherence
length is larger than 0.55 times the width of the slab, then vortices cannot be
present in the slab, whereas for smaller coherence lengths they can. This change
of regime is accompanied by a change of behavior of the temperature--field
dependence along the normal--superconducting (N-S) phase boundary. The second derivative
of the transition temperature with respect to the field is discontinuous at the critical
value of the coherence length. These results were confirmed experimentally.\cite{crossov}
For more recent studies on slabs, see e.g. Refs.~\onlinecite{Schuller,Bariloche,VB}.

We consider a ring with perfect axial symmetry and analyze it by means of the
Ginzburg--Landau model. Since the magnetic field varies with time, it is natural
to use the time-dependent Ginzburg--Landau model (TDGL). This approach has been
followed in recent calculations.\cite{vodo} While TDGL is appropriate for the
study of a given experiment, the number of parameters it introduces is rather
large, depending on the material, the temperature, and the rate at which the 
magnetic field is swept. In this article we have chosen to consider the theoretical
limit in which the field varies ideally slowly, so that the time independent 
formalism can be used.

We assume that the ring is very thin compared with the
magnetic penetration length. Under these assumptions the free energy is
(up to unimportant constants)
\begin{equation}
G=\int_0^{2\pi}\!\!\!d\theta\int_{R_{\rm i}}^R\!\!r\,dr \left[\mu(-|\psi|^2+
\frac{1}{2}|\psi|^4)
+|(iR\bm{\nabla}-b r/R\hat\theta)\psi|^2\right]
\label{GGL}
\end{equation}
where $r$ and $\theta$ are cylindrical coordinates, $\psi$ is a normalized 
order parameter, $R_{\rm i}$ and $R$ are
the inner and outer radii of the ring, $\mu=(R/\xi)^2$ is a function of the 
temperature ($\xi$ is the coherence length) and $b=\pi R^2 H_{\rm e} /\Phi_0$, 
where $H_{\rm e}$ is the magnetic field and $\Phi_0$ the quantum of flux. 

\section{\label{onset}THE ONSET OF SUPERCONDUCTIVITY}
In order to gain some intuition about this problem, we first consider the
Little--Parks regime, at the onset of superconductivity.
In this regime $|\psi|\ll 1$, so that the Ginzburg--Landau equations are linearized
and the case of the ring can be solved exactly.\cite{James,Zwerger,Bruind}
The purpose of the present section is to point out that 
there exists a critical point in the N-S boundary. As will
be seen in the following sections, it appears that this critical point plays 
a major role in the ``superheating" stability of the superconducting states.

\begin{figure}
\scalebox{0.85}{\includegraphics{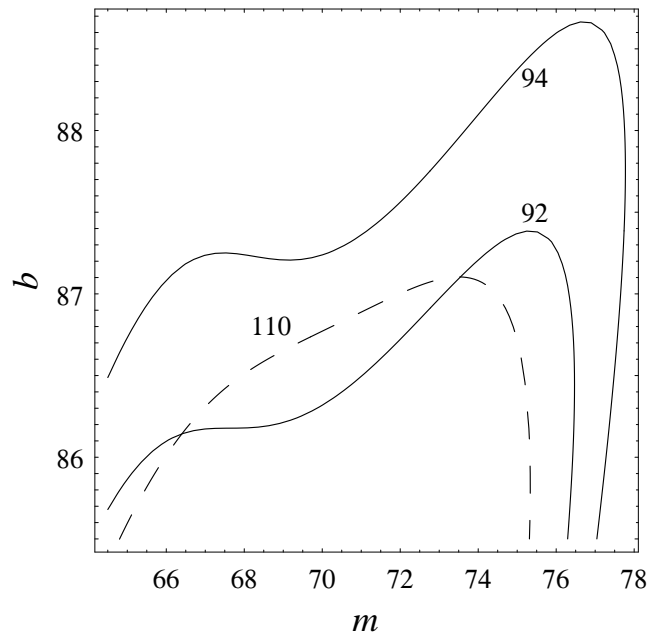}}%
\caption{\label{sub-hyp}The continuous lines mark the onset of superconductivity
in the flux versus number of fluxoids plane. The ring is superconducting 
in the area ``inside" 
the appropriate line. Only integer values of $m$ are meaningful. The ratio between
the inner and the outer radius is 0.8. The lower line is for
$\mu=92$ and the upper line, for $\mu=94$. The dashed line is a line of constant 
energy for $\mu=110$. The continuous lines were obtained by an exact calculation;
the evaluation of the dashed line approximates the radial dependence of the order
parameter by a sum of three orthogonal polynomials.}
\end{figure}

Figure~\ref{sub-hyp} shows the normalized field $b$ at which a ring with 
$R_{\rm i}=0.8R$
becomes superconducting, as a function of the winding number $m$, for $\mu=92$
and for $\mu=94$. Only integer values of $m$ are meaningful, and only the 
value of $m$ with the highest $b$ is physically realized, but these curves are
useful for explanatory purposes. We found that for $\mu\alt 92$ these onset curves
have only one maximum, whereas for $\mu\agt 94$ they have two. At 
$\mu\approx 93,m\approx 68,b\approx 87$ there is a critical point, where the 
first and the second derivative of the onset field (or of the free energy)
with respect to $m$, both vanish. As a consequence, near this critical point,
the free energy depends very weakly on the winding number $m$.

The influence of this critical point is felt
quite far from its immediate vicinity. The dashed line in Fig.~\ref{sub-hyp} is a
line of constant (negative) free energy for $\mu=110$. We see in its upper part 
that for a range of about 8 winding numbers the slope of $b$ with respect to $m$,
expressed in ``natural" units, is less than a tenth. By the relationship
$(\partial b/\partial m)_G=-(\partial G/\partial m)_b/(\partial G/\partial b)_m$, 
the smallness of $(\partial b/\partial m)_G$
suggests that the free energy also has a very weak dependence on the winding 
number. 

In reality, only integer values of $m$ and their linear combinations exist. 
The typical scenario for flux transitions is as follows. A given field $b$ is
present and the order parameter has a certain winding number $m$. Linear combinations
with other winding numbers act as energy barriers and the sample is ``trapped"
at $m$. When $b$ becames sufficiently small, the energy barrier towards smaller
winding numbers is levelled off, and the state decays. If $(\partial G/\partial m)_b$
is large, the order parameter will be ``stopped" by the following barrier and
become trapped at $m-1$.
However, a small slope means that $m$ has to change by a big amount
for a given change in $G$, leading to the possibility
of transitions that change $m$ by several units.

The critical point found here can be associated to two phenomena which have been 
discussed in the literature. The first is the condition for the formation of
vortices in a slab, quoted in Sec.~\ref{INT}. Using the value $0.2R$ 
as the thickness of the slab, the onset for the appearance of vortices is located at
$\mu\sim b\sim 80$, not far from the critical point in our case. We can
therefore expect that for $\mu$ above the critical point (low temperatures) there
will be vortices in the ring, whereas for small $\mu$ the winding number will be the 
same for the inner or for the outer boundary of the ring.

An additional feature noted in Ref.~\onlinecite{book} is the crossover from a regime
in which at the N-S boundary $b$ is approximately proportional to $\mu^{1/2}$ to a regime 
where $b$ is approximately proportional to $\mu$. In 
the case of rings, this crossover applies to the background line of this boundary,
which has the Little-Parks oscillations superimposed on it.\cite{Bruind,JR} This is
actually a dimensional crossover: at low fields the sample behaves as a thin slab
and $b=5(3\mu)^{1/2}$; at high fields it behaves as half a plane and $b=0.847\mu$.
Again, the crossover occurs at $\mu\sim 100$. For the case of slabs the passage from
one regime to the other involves discontinuities of some quantities, but for the case
of rings these discontinuities are not relevant, since except for accidental situations
the critical point does not occur at a physical (i.e. integer) value of $m$.

\begin{figure}
\scalebox{0.85}{\includegraphics{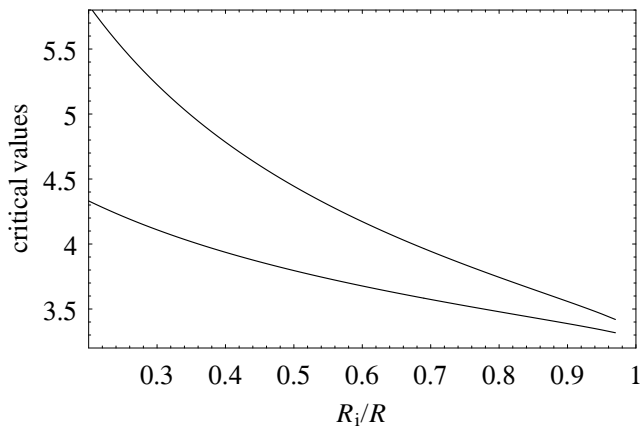}}%
\caption{\label{critval}Normalized values of $\mu$ and $b$ at which the critical 
point occurs, as functions of the ratio between the inner and the outer radius 
of the ring. The upper curve is $\mu(1-R_{\rm i}/R)^2$ and the lower curve is 
$b(1-R_{\rm i}/R)^2$.
In the calculation of these curves we have approximated the radial dependence of 
the order parameter by a sum of three orthogonal polynomials.
For $R_{\rm i}\rightarrow R$, the critical values of $\mu$ and $b$ diverge and our
method cannot be used; in this limit, the ring becomes a slab and we expect
$\mu(1-R_{\rm i}/R)^2\rightarrow 3.39$ \cite{Fink1} and 
$b(1-R_{\rm i}/R)^2\rightarrow 3.24$.\cite{Schul} }
\end{figure}

Since we believe that the main features of the situation we study are governed by
the critical point described in this section, we have evaluated numerically the
position of this point as a function of the ratio $R_{\rm i}/R$. 
Figure \ref{critval} shows the values of $\mu$ and $b$ at which the critical point
occurs. Since to a rough approximation these 
values are inversely proportional to the square of the width of the ring, the
normalized values $\mu(1-R_{\rm i}/R)^2$ and $b(1-R_{\rm i}/R)^2$ appear in the figure. 
In the following sections we will always take $R_{\rm i}=0.8R$,
which is a representative experimental value.

\section{MATHEMATICAL METHOD}
We return now to the general situation in which the order parameter $\psi$ is
not necessarily small.
We write $\psi$ as a function of the cylindric coordinates
$r$ and $\theta$ and assume that the ring is so thin that there is no $z$-dependence.
We expand
\begin{subequations}
\label{expansion}
\begin{equation}
\psi(r,\theta)=\sum{\mathcal R}_m(r) e^{-mi\theta} \;,
\label{expthet}
\end{equation}
\begin{equation}
{\mathcal R}_m(r)=\sum c_{mn}P_n(r) \;,
\label{exp-r}
\end{equation}
\end{subequations}
where Eq.~(\ref{expthet}) is a Fourier expansion and Eq.~(\ref{exp-r}) is an expansion
into orthogonal polynomials, with $P'_n(R_{\rm i})=P'_n(R)=0$ and such that $P_n(r)$ 
vanishes $n$ times in the range $R_{\rm i}\le r\le R$. We then keep up to four terms in 
Eq.~(\ref{expthet}): two ``leading" terms with winding numbers $m_0$ and $m_1$,
and two ``satellite" terms, with winding numbers $2m_0-m_1$ and $2m_1-m_0$.
In Eq.~(\ref{exp-r}) we keep the terms $n=0,1,2$ for the leading terms, and only
the term $n=0$ for the satellite terms. The coefficients $c_{mn}$ which are not declared
to vanish, are found by minimizing the free energy. 

The physical meaning of a single winding number in the expansion (\ref{expthet})
is that $|\psi|$ has axial symmetry. If there are two leading winding numbers $m_0$ 
and $m_1$ and two satellites, as we shall consider as our most general situation,
and if the weights of the two leading winding numbers are similar, then the equilibrium
order parameter exhibits a ring of $|m_0-m_1|$ equidistant vortices in the
superconducting sample. 
If there are several winding numbers, but the weight of one of them is much larger
than those of the others, then $|\psi|$ does not have axial symmetry, but there are
no vortices in the sample.
For situations in which there are vortices in the sample 
that are not equidistant from the $z$-axis, our approximation breaks down.

\begin{figure}
\scalebox{0.85}{\includegraphics{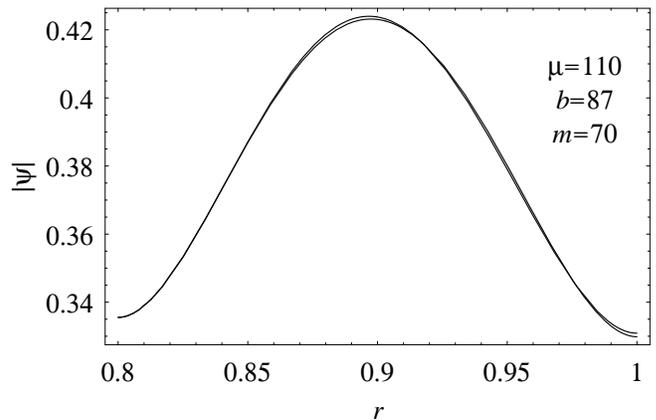}}%
\caption{\label{check}Comparison of approximated and exact values of the order
parameter for a typical situation. The difference between both curves does not
exceed $0.001$.}
\end{figure}

According to our results, the approximations used
seem to be reasonable as long as $\mu$ is not much higher than that of the critical
point discussed in Sec.~\ref{onset}. The usual situation is that the weight of the
leading terms is much greater than that of the satellites. In Fig.~\ref{check} we compare
the radial dependence of an axially symmetric order parameter obtained by the present
approximations to that from an exact calculation.

Our procedure is as follows. For fixed temperature (fixed $\mu$), we first evaluate the
magnetic field for the onset of superconductivity. At the onset, the Ginzburg--Landau
equations are linear, so that the order parameter generically has axial symmetry and the sum
(\ref{expthet}) contains only one $m$, which we denote by $m_0$. The field is then
gradually lowered and for every field we evaluate the order parameter 
${\mathcal R}_{m_0}(r) e^{-{m_0}i\theta}$. We then check the stability of this 
order parameter against small variations.

Variations explore subspaces with larger dimensions than that of the equilibrium
order parameter. When looking for decays from $m_0$, variations
also contain the winding numbers $m_0-\ell$
and $m_0+\ell$, where $\ell$ is swept over at least the range $1\le\ell\le\mu/5$. 
It is important to include both $m_0-\ell$ and $m_0+\ell$, since due to nonlinearity
of the GL model these harmonics are coupled. For 
each variation we evaluate the Hessian of the variation of the free energy and obtain 
its eigenvalues. If all the eigenvalues are positive, it means that 
${\mathcal R}_{m_0}(r) e^{-{m_0}i\theta}$ is stable, and we proceed to decrease the
field by a small step.

If any of the eigenvalues is negative, it means that 
${\mathcal R}_{m_0}(r) e^{-{m_0}i\theta}$ is unstable
with respect to the variation represented by the corresponding eigenstate. 
We denote by $m_1=m_0\pm\ell$ the value of the
winding number that has the highest weight in the variation against which $m_0$
was unstable. 

During the decay of ${\mathcal R}_{m_0}(r) e^{-{m_0}i\theta}$, the order parameter
changes quickly. Therefore, in order to follow the order parameter or the currents
during the decay, we would have to use the TDGL model. Using a gauge in which the
electrochemical potential vanishes, the equations of this model take the form
\begin{subequations}
\label{TDGL}
\begin{equation}
\partial \psi/\partial t=-D_\psi \delta_{\bar\psi} G \;,
\end{equation}
\begin{equation}
\partial {\bf A}/\partial t=-D_{\bf A} \delta_{\bf A} G \;,
\end{equation}
\end{subequations}
where ${\bf A}$ is the vector potential, $\delta_{\bar\psi}$ and
$\delta_{\bf A}$ denote variations over the complex conjugate of $\psi$ and over
${\bf A}$, and $D_\psi$ and $D_{\bf A}$ are positive constants. From here, it
follows that $G$ will always decrease during the decay and the order parameter
will reach equilibrium when $G$ reaches a minimum. (The applied field is assumed
to remain constant during the decay.)

Therefore, in order to obtain the order parameter after the decay,
we perform a minimization of the free energy over states of
the form (\ref{expansion}), where the leading winding numbers are $m_0$ and $m_1$.
The initial state from which the minimization flows is a combination of the former
state with $m=m_0$ and of the eigenstate that had the negative eigenvalue.
The minimization may lead to a state in which only one winding number has a
significant weight; in this case we recover the previous situation and we just
have to update the value of $m_0$. The other possibility is to be left with
a combination of winding numbers (which means that $|\psi|$ is not axially symmetric).
It seems plausible to assume that in the generic case the minimization leads to
the same minimum, and with the same leading harmonics that would be obtained by
TDGL; we intend to investigate the range of validity of this assumption in 
a future study.

In any numerical procedure, arbitrary decisions have to be taken concerning
parameters such as cutoffs and convergence factors. Usually these parameters are
fixed empirically in order to optimize convergence properties.
In our case, one has to choose the initial weights given to the 
original state and to the deviation that renders it unstable
in their initial combination, the minimization method,
and the minimum weight below which a winding number is considered ``not significant".
We have used several criteria and become
convinced that, in most cases, our results do not depend on the particular choices.

If the order parameter is of the form (\ref{expthet}) with leading winding numbers
$m_0$ and $m_1$, we check its stability along similar lines to those described for
the axiallly symmetric case. This time the attempted variations contain combinations of 
winding numbers $\ell$, $2m_0-\ell$ and $2m_1-\ell$, where $\ell$ is swept over at least 
the range $[\max(m_0,m_1)-\mu/5]\le\ell\le[\max(m_0,m_1)-1]$.

\section{RESULTS}

For each field, after the order parameter is known, the free energy is evaluated
using Eq.~(\ref{GGL}), the normalized current density
is evaluated as ${\rm Re}[\bar\psi(iR\bm{\nabla}-b r/R\hat\theta)\psi]$ and then
the current (per unit height) is obtained by integration over $r$ of the 
tangential component.

\subsection{Near $T_c$}

\begin{figure}
\scalebox{0.85}{\includegraphics{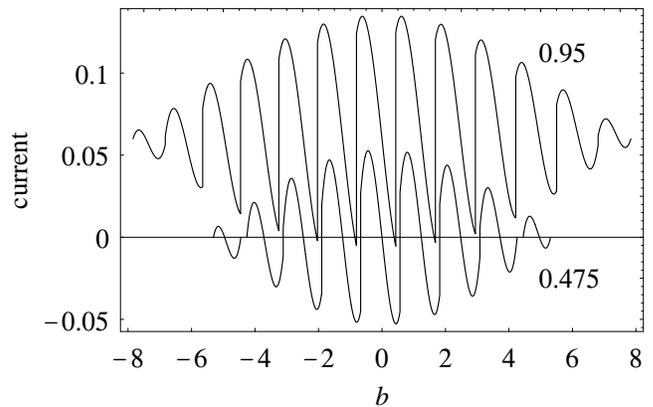}}%
\caption{\label{lowmu}Normalized current as a function of the normalized field for 
$\mu=0.475$ and for $\mu=0.95$. Decays are represented by the vertical lines at the
left side of each undulation. For visibility, the curve for $\mu=0.95$ has been
raised by $0.06$. The physical current per unit height, in electrostatic units, is
obtained by multiplying the normalized current by $c\Phi_0\mu/2(2\pi\kappa R)^2$,
where $c$ and $\kappa$ are the speed of light and the Ginzburg--Landau parameter.}
\end{figure}

Figure \ref{lowmu} shows the current around the ring, as a function of the field,
for low values of $\mu$. As expected, there are discontinuities in the current,
nearly periodically spaced. These discontinuities correspond to the passage between
consecutive winding numbers.

For $\mu=0.475$ and close to $b=\pm 4.35$, the ring is in
the normal state. This reentrant behavior is due to the Little--Parks oscillations.
For a ring with $R_{\rm i}=0.8R$, reentrant superconductivity is possible in the
range $0.02\le\mu\le 12.4$. For $\mu<0.02$, only the Meissner state ($m=0$) exists;
in the range $0.02\le\mu\le 12.4$ there are ``windows" for which superconductivity
is interrupted when passing from one $m$ to the next. For low values of $\mu$ these
windows are wide, but, as $\mu$ increases, the background slope of the N-S boundary
in the $b-\mu$ plane increases and these windows become narrower. For $\mu>12.4$
the N-S boundary is monotonic.

\subsection{Short coherence length}
\begin{figure}
\scalebox{0.85}{\includegraphics{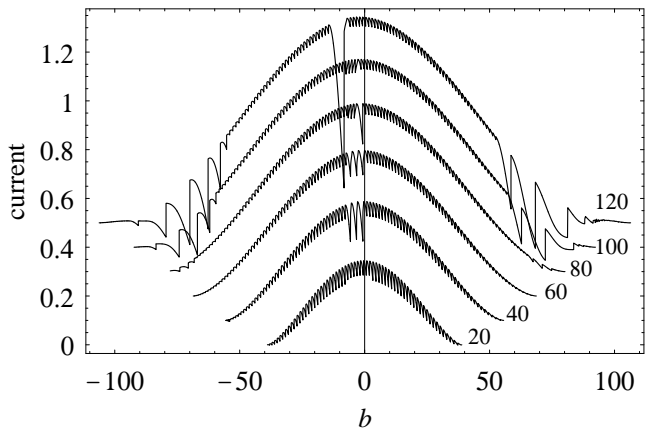}}%
\caption{\label{curr}Normalized current as a function of the normalized field for 
a wide range of values of $\mu$. The value of $\mu$ is written next to each curve.
Shifts of $0.1$ have been inserted between consecutive lines. The right end
of this graph is shown enlarged in Fig.~\ref{highI}.}
\end{figure}

\begin{figure}
\scalebox{0.85}{\includegraphics{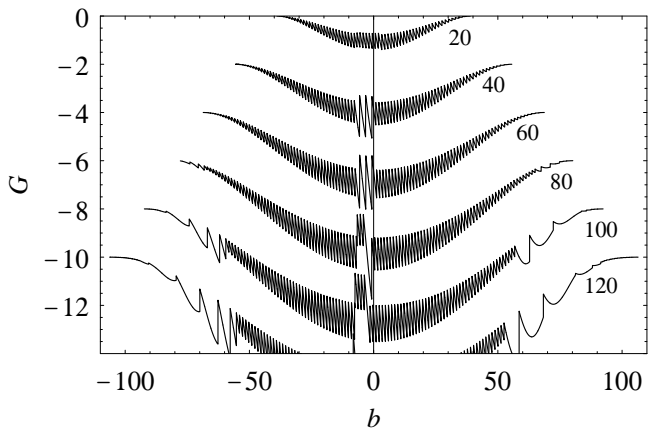}}%
\caption{\label{G}Free energy as a function of the normalized field for
the same values of $\mu$ as in Fig.~\ref{curr}.
Shifts of $-2$ have been inserted between
consecutive lines. The central part of $\mu=120$ has been chopped. The right end
of this graph is shown enlarged in Fig.~\ref{highG}.}
\end{figure}

Figures \ref{curr} and \ref{G} show our results in the range $20\le\mu\le 120$.
For $\mu>120$, our approximation of Eq.~(\ref{expthet}) by four uniformly spaced
harmonics breaks down. This was expected since, if the distance between the 
boundaries is considerably larger than the coherence length, then there is room
for more than one ring of vortices.

Obviously, if the magnetic field is reversed, the current in the sample for
the equilibrium (absolute minimum) configuration is reversed too. For small values 
of $\mu$ hysteresis is a minor effect, and indeed we note in Fig.~\ref{lowmu} that the 
current for $\mu=0.475$ is nearly an odd function of the field. On the other
hand, for the values considered in Fig.~\ref{curr}, hysteresis is dominant
and the equilibrium values have practically no importance. At first glance
the curves in Fig.~\ref{curr} look like even functions and we observe
that qualitatively different regions appear in nearly
symmetric pairs. In the following, we shall review these regions, by decreasing
values of $|b|$. Following the line of reasoning of Sec.~\ref{onset}, we may argue
that the approximate evenness is due to the invariance of $|(\partial G/\partial m)_b|$
when both $b$ and $m$ change sign.

\subsubsection{Near the onset}
\begin{figure}
\scalebox{0.85}{\includegraphics{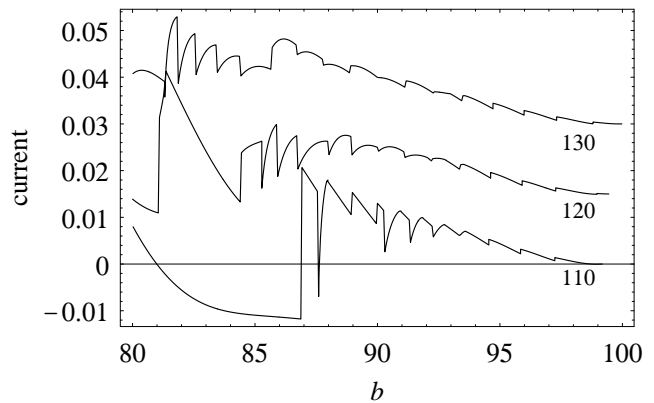}}%
\caption{\label{highI}Normalized current near the onset of superconductivity.
The lowest line is for $\mu=110$ and sits at its true position. The other lines
are for $\mu=120$ and $\mu=130$. For each increment of $\mu$ by 10, the position
of the line is raised by 0.015 and is also shifted by 7 to the left.}
\end{figure}

\begin{figure}
\scalebox{0.85}{\includegraphics{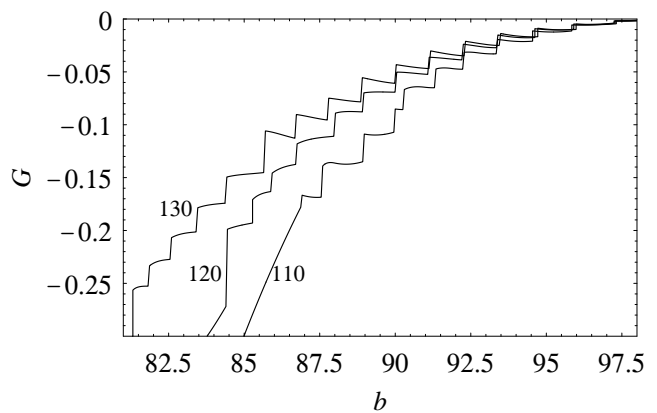}}%
\caption{\label{highG}Free energy near the onset of superconductivity.
The steepest line is for $\mu=110$ and sits at its true position. The other lines
are for $\mu=120$ and $\mu=130$. For each increment of $\mu$ by 10, the position
of the line is shifted by 7 to the left.} 
\end{figure}

Figures \ref{highI} and \ref{highG} show our results for high values of $b$,
with $\mu=110$, $120$ or $130$. [Near the onset of superconductivity, the
approximation of the expansion (\ref{expthet}) by a limited number of terms
remains valid beyond $\mu=120$.]
The region at the right consists of steps where a single value of $m$ is left
in (\ref{expthet}) and, as $b$ is lowered, $m$ decreases by 1 between consecutive
steps. There are five such steps for $\mu=120$ and $130$, and four steps for 
$\mu=110$. The current always drops when the state decays to a lower $m$.

\subsubsection{Breaking of axial symmetry}
This is roughly the region $87\alt b\alt 94$ for $\mu=110$, 
$81+7\alt b\alt 94+7$ for $\mu=120$ and includes most of the line for $\mu=130$
in Fig.~\ref{highI}. If we denote
by $(m_0)$ a state with a single harmonic $m=m_0$ and by $(m_0,m_1)$ a state 
with two leading winding numbers $m_0$ and $m_1$, this region extends from
$(84,70)$ to $(77,65)$ for $\mu=110$, from $(91,74)$ to $(78,66)$ for $\mu=120$ 
and from $(98,79)$ to an uncalculated state for $\mu=130$. 

We may divide this region into a subregion of ``incipient
symmetry breaking" and a subregion of ``developed symmetry breaking". 
The subregion of incipient
symmetry breaking consists of the sequence of double steps 
$(m_0^{(1)})\rightarrow(m_0^{(1)},m_1^{(1)})\rightarrow(m_0^{(2)})\rightarrow(m_0^{(2)},
m_1^{(2)})\rightarrow\cdots$. Here 
$m_0^{(i+1)}=m_0^{(i)}-1>m_1^{(i)}$, so that the transitions 
$(m_0^{(i)},m_1^{(i)})\rightarrow(m_0^{(i+1)})$ are discontinuous; on the other
hand, the transitions $(m_0^{(i)})\rightarrow(m_0^{(i)},m_1^{(i)})$ are continuous
bifurcations. In the subregion of developed symmetry breaking, all the states are
combinations with two leading winding numbers. For $\mu=130$ there is only one
step with incipient symmetry breaking and for $\mu=120$ there are two steps; for
$\mu=110$, incipient symmetry breaking extends over the entire region. Some segments
are very short, and cannot be seen in the graph.

The subregion of developed symmetry breaking can be subdivided further: in its 
high-field part the transitions 
$(m_0^{(i)},m_1^{(i)})\rightarrow(m_0^{(i+1)},m_1^{(i+1)})$ have either 
$m_0^{(i+1)}=m_0^{(i)}$ or $m_1^{(i+1)}=m_1^{(i)}$; for lower fields there may be
``cascades". By a cascade we mean that, after a state decays, the new state is
already unstable with respect to a third state, and decays at the same field at
which it appeared. By means of cascades, $m_0$ and $m_1$ can both change simultaneously.

For axially symmetric states, the slope of the current in Fig.~\ref{highI} is negative.
On the other hand, for combinations with two leading winding numbers, the current is
a markedly convex function, and can reach large positive slopes. For decays that follow 
after these large positive slopes, the current usually {\em rises}.

\subsubsection{Critical region}
This is the region with large discontinuities near $|b|\sim 70$. For $\mu=120$ and
$\mu=110$
the region of asymmetry ends with a long cascade, the last stage of which is
a discontinuous decay into a symmetric state. For $\mu\le 100$ the region of 
asymmetry is absent. For $\mu=100$ the critical region begins with a decay in which $m$ 
overshoots and then goes up. For $b<0$, the critical region in which $m$ jumps by
several units appears without previous signs. 

It may be surprising that although the samples are sufficiently wide to 
have vortices on them, the order parameter is axially symmetric in this region.
An intuitive explanation might be that the field is not sufficiently large
and therefore a ring of vortices is not thermodynamically favorable.

Table~\ref{states} is a record of
the states through which the critical region passes. 

\begin{table}
\caption{\label{states}Fluxoid numbers occurring in the critical region. Negative
numbers are for the region in the negative field.}
\begin{ruledtabular}
\begin{tabular}{cc}
$\mu$ & states \\
\hline
120 & $(61)\rightarrow (53)\rightarrow (48)$ \\
120 & $(-39)\rightarrow (-41)\rightarrow (-45)\rightarrow$~~~~~~~~~~~~~~~~~~\\
& ~~~~~~~~~~~~~~~~~~$(-51)\rightarrow (-59)\rightarrow (-69)\rightarrow (-79)$ \\
110 & $(66)\rightarrow (57)\rightarrow (51)\rightarrow (48)$ \\
100 & $(74)\rightarrow (64)\rightarrow (56)\rightarrow (51)$ \\
100 & $(-43)\rightarrow (-45)\rightarrow (-49)\rightarrow (-55)
   \rightarrow (-63)\rightarrow (-72)$ \\
95 &  $(75)\rightarrow (65)\rightarrow (58)\rightarrow (53)\rightarrow (51)$ \\
90 &  $(72)\rightarrow (64)\rightarrow (57)\rightarrow (53)$ \\
80 &  $(65)\rightarrow(62)\rightarrow (59)\rightarrow (57)$ \\
80 &  $(-51)\rightarrow(-53)\rightarrow(-56)\rightarrow(-60)\rightarrow(-64)$
\end{tabular}
\end{ruledtabular}
\end{table}

One notices that for $b>0$ the
number of fluxoids that go through the ring during the critical region is largest
for $\mu$ near its critical value. The largest signal in the 
current--field graph, as well as the number of fluxoids that go through for $b<0$,
appear to be at higher values of $\mu$.

\subsubsection{Subcritical region}
This is the broadest region in Figs. \ref{curr} and \ref{G}. As the field decreases
through the critical region, jumps in the winding number decrease from giant to
moderate, until we finally reach a region in which the winding number changes by just
one in each decay. For $\mu\le 60$ the two previous regions are absent and there
is no distinction between this and the first region.

For $20\le\mu\le 60$ and $b>0$, the average range $\Delta b$ between consecutive
jumps in this region is $1.30$; for $b<0$, $\Delta b=1.17$. This difference can be understood,
since for $b>0$ a delay in the decays is produced by ``supercooling" and in the
negative region it is necessary to catch up with this delay. For $80\le\mu\le 120$,
$\Delta b=1.24$ for positive fields and $\Delta b=1.23$ for $b<0$, indicating that
the presence of the critical region helps to release most of this delay. Note that
$\Delta b$ is not close to unity due to our choice of the outer radius as the unit
of length. Had we chosen the average radius $0.9R$ as the unit of length (as in 
Ref.~\onlinecite{Price}), we would have had to divide the values of $b$ by 
$1/0.9^2\approx 1.235$.

\subsubsection{Low fields}
This is a short region centered at slightly negative fields. In this region the
``satellite" harmonic $2m_1-m_0$ is not necessarily small. In this region the
results we obtain are sensitive to the minimization strategy, and the results
in Figs. \ref{curr} and \ref{G} are less reliable here than in the other regions. 

\section{COMPARISON WITH EXPERIMENTS}
\begin{figure}
\scalebox{0.85}{\includegraphics{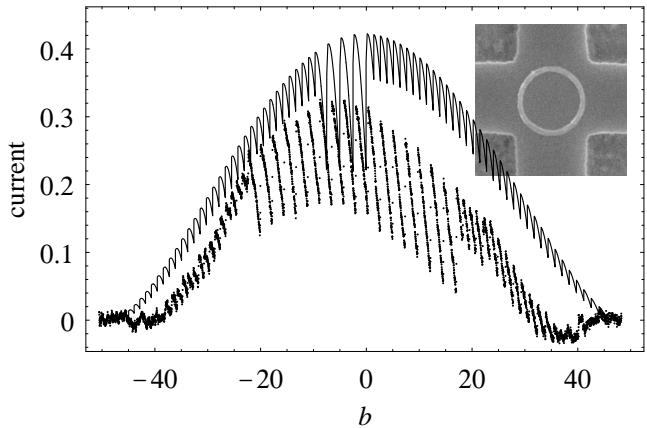}}%
\caption{\label{data}
Comparison between our results and those measured in Ref.~\onlinecite{GM}.
The experimental data are represented by the dense set of dots and the theory
by the line above.
The experimental value of $b$ was obtained by multiplying the applied field by
$\pi R^2/\Phi_0$. Also, since in our calculations the field was swept downwards
and in the experiment upwards, the sign of the experimental field was switched.
The experimental value of the normalized current was obtained by multiplying
the field induced by the supercurrent at the center of the ring by
$2\pi\kappa^2(R_{\rm i}+R)R^2/\Phi_0 \mu t$. We took $\kappa\sim 0.5$ and estimated
the field induced at the center of the ring as 2.5 times its average value
on the active square of the Hall cross. The calculated curve is for $\mu=30$.}
\end{figure}

To our knowledge, the experiments closest to the situation discussed in this 
article are Ref.~\onlinecite{Price} and Refs.~\onlinecite{Ped,GM}. Reference 
\onlinecite{Price} deals with only three fluxoids, and from the present
point of view may be considered featureless. (See however Ref.~\onlinecite{P2}.)
The results of Ref.~\onlinecite{GM} are similar to those of Ref.~\onlinecite{Ped},
but in this case the susceptibility goes to zero at the onset of superconductivity,
as in our results. Ref.~\onlinecite{GM} also confirms our result that, if the
width of the ring is appreciably smaller than the coherence length, all the 
transitions involve the passage of just one fluxoid.

Reference \onlinecite{Ped} reports on two rings. In both cases the estimated
mean radius is $2.16\mu$m and the coherence length, 180nm. For their wide
ring, $R-R_{\rm i}=630$nm. Using graph~\ref{critval}, these values result in
$\mu=(R/\xi)^2\sim 3\mu_{\rm crit}$, far beyond the region we have studied.
One observation can nevertheless be adventured: the critical value of the
field is $b_{\rm crit}\sim 56$, which for this geometry corresponds to about
60G. The ``catastrophic behavior" in their Fig.~5 begins not far from there.

Their narrow ring has a width $R-R_{\rm i}=316$nm, which now implies
$\mu=(R/\xi)^2\sim 0.9\mu_{\rm crit}$. Translated to our case, this corresponds
to $\mu\sim 80$. Inspection of their Fig.~2 and our Fig.~\ref{curr}
suggests that $\mu=60$ would give a better comparison. The difference between
$\mu=80$ and $\mu=60$ could be due to a 15\% error in any of the reported lengths,
and is an expected experimental uncertainty. There are some similar features
between our results and the experimental curve: there is a central region 
sandwiched between two outer regions, like the low-field and the
subcritical regions of the previous section. The size of the discontinuities
grows towards lower fields at an apparently correct pace. The average period
$\Delta b$ for positive fields is longer than that for negative fields, and the 
ratio $1.30:1.17$ seems to be of the right order.
For $\mu=60$ and $b\sim 0$ we predict a current of the order of 
$0.6c\Phi_0\mu t/2(2\pi\kappa R)^2$, where $t$ is the thickness of the ring.
Assuming $t\approx 90$nm and $\kappa\sim 0.5$, this current amounts to 
$2\times 10^6$esu/sec. A current like this would induce a magnetic field of 
$\sim 2$G at the center of the ring; this is the correct order of magnitude.

Figure \ref{data} compares our results with a measurement in Ref.~\onlinecite{GM}.
The micrograph at a corner shows an aluminum ring grown on a Hall cross which permits
to measure the average magnetic field produced by the supercurrent.\cite{Geim} 
The dots are the experimental results. As can be
seen from the micrograph, the width is close to $0.2R$, enabling a direct
comparison with our results. The thickness is $t\approx 0.12\mu$m. The outer
radius, as measured by optical means, is $R=1.1\pm 0.1\mu$m. We picked
$R=1.22\mu$m, in order to have the correct average periodicity. The temperature
was $0.6\pm 0.1$K, which implies a coherence length between $0.196\mu$m and
$0.217\mu$m and hence $\mu=34\pm 3$. The line in Fig.~\ref{data} was evaluated for
$\mu=30$. The difference in height between the measured and the calculated results
is not very significant, since several factors have been just estimated; 
however, the range of the central part is much wider than predicted. Again,
the average period $\Delta b$ for positive fields is longer than that for negative 
fields; this time the effect is more pronounced than in Ref.~\onlinecite{Ped}.

Since both experiments show a central part with two-fluxoid jumps which is 
considerably wider than in our results, we list some deviations from our 
idealizations that might give rise to this discrepancy.
\begin{itemize}
\item We have neglected the induced field. The magnetic penetration depth
$\lambda=\kappa R\mu^{-1/2}\sim 0.5\times 2300{\rm nm}\times 60^{-1/2}\approx 150$nm
is only marginally larger than the thickness of the ring, so that our assumption 
is not clearly justified. 
The influence of the induced field on ``superheating" for
the case of thick disks has been considered in Ref.~\onlinecite{PhC} and does not 
hint at a qualitatively different behavior.
\item An experimental ring never has perfect axial symmetry. This imperfection 
has been treated in previous studies\cite{JB,Spring} and has a profound influence
near the onset of superconductivity, but is not expected to be important deep
inside the superconducting region. Large imperfections have been found to inhibit
the size of the transitions.\cite{GM,VB}
In the present problem, variation of the width
with the height might also be important.
\item Decays most probably occur before the ideal metastability limit is reached,
mainly due to electromagnetic noise. When taking measurements, noise can be filtered 
out by a lock-in amplifier, but in order to avoid premature decays only shielding 
could help.
\end{itemize}

\begin{acknowledgments}
Special thanks to A. Geim for sending me the data and the micrograph in Fig.~\ref{data}
prior to publication. This article has benefited from comments raised by 
O.~Megged, S.~Pedersen and J.~Rubinstein.
\end{acknowledgments}
\bibliography{Linde5}
\end{document}